\documentclass[prc, reprint, amsmath, amssymb, superscriptaddress, nofootinbib, showpacs]{revtex4-1}

\usepackage{graphicx}%
\usepackage{multirow}
\usepackage{float}
\usepackage{epstopdf}
\usepackage[utf8]{inputenc}
\usepackage[english]{babel}
\usepackage{footnote}
\usepackage[justification=justified,
   format=plain]{caption}
\usepackage{xcolor}
\usepackage{textcomp}

\usepackage[skip=6pt plus1pt, indent=12pt]{parskip}
\begin{document}

\title{Undergraduate physics students’ experiences: Exploring the impact of underrepresented identities and intersectionality}%

\author{D. K. Keblbeck}
\affiliation{Department of Physics, Colorado School of Mines, Golden, CO 80401, USA}

\author{K. Piatek-Jimenez}
\affiliation{Department of Mathematics, Central Michigan University, Mount Pleasant, MI 48859, USA}

\author{C. Medina Medina}
\affiliation{Dematic, KION Group, Holland, Michigan 49423, USA}

\date{\today}%

\begin{abstract}
Historically, physics has been a predominantly male field, with previous literature showing that there is little diversity amongst U.S. physics students at the undergraduate and graduate levels, or amongst physicists within the work force. Recent research indicates that the lack of diversity in physics is partially due to an unwelcoming climate within physics departments, as well as differential experiences during college. Most physics education research that addresses the lack of underrepresented identities within the field has focused on the identities of women and people of color. There has been little research to investigate people with multiple underrepresented identities, including those such as socioeconomic status, first-generation college students, or learning disabilities. Furthermore, there has been even less research conducted to better understand the impacts of the intersection of these underrepresented identities and how it relates to experiences when pursuing a physics major. In order to address this gap in the literature, our research project has investigated undergraduate physics students' experiences, to better understand what factors affect their experiences and how these may differ by the intersection of one's underrepresented identities. In particular, we explored how these identities impact their experiences as a physics major. To achieve this goal, we conducted a series of in-depth interviews with physics majors at one university to learn more about their college experiences regarding physics. Our findings suggest that there is a disproportionate number of obstacles when pursuing a physics major faced by those with a greater number of underrepresented identities. We conclude that there is a need for more equitable pedagogical practices and departmental policies within the undergraduate physics experience, in addition to a more ``human" approach to mentorship, in order to foster an environment in which students with underrepresented identities can feel supported and thrive academically and professionally.
\end{abstract}

\maketitle

\section{Introduction and Background}
Women and people of color have been historically underrepresented in the fields of Science, Technology, Engineering, and Mathematics (STEM). For instance, in 2020 only 38$\%$ of bachelor’s degrees earned in STEM in the United States were awarded to women and in the same year only 26$\%$ of these degrees were awarded to people of color \cite{aps_women_stem, nsf_minorities}.

This underrepresentation of women and certain racial groups in the United States is even more bleak in the subfield of physics, with only 25$\%$ of women and 16$\%$ of students of color being awarded a physics bachelor’s degree in 2020 \cite{aps_minority_physics}.
The literature suggests many reasons that contribute to this underrepresentation. In particular, one contributing factor appears to be the culture within certain STEM disciplines, including physics. In these traditionally male-dominated fields, for instance, women often face cultural beliefs that men are better at, or more suited for, these disciplines \cite{cech2011}. Moreover, previous research suggests that members of marginalized groups who experience unwelcoming college cultures tend to experience social and academic withdrawal and isolation \cite{yosso2009, hurtado1998, ali2006, strayhorn2010}. 
Within the field of physics specifically, previous research has documented a cold and discouraging climate within physics departments towards women and members of certain racial groups \cite{barthelemy2016, doucette2020, abdurrahman2021}. Prior research indicates this can be a result of an unawareness within the culture of physics toward the differential experiences that members of marginalized groups face \cite{santana2021}, leading to inequitable treatment. Some college students have reported experiencing a culture of arrogance around being a physicist, which has been attributed to why some women leave their physics major \cite{bremer2017}. Another study reported female undergraduate physics majors experiencing an unpleasant and exclusionary culture created by their male peers and instructors, creating feelings of isolation \cite{santana2023}. 

Another suggested factor in the lower selection and retention rates of people with underrepresented identities in physics is the lack of support they receive to develop their physics identity \cite{hazari2010}, or how they view themselves as a physics person. One research study indicated that the compounding effects of systemic inequity faced by people with intersecting minoritized identities relating to race/ethnicity, gender, and/or sexuality can hinder the development of a physics identity and lead to a decline in academic performance in physics \cite{quichocho2020}. Additionally, not feeling recognized as a physics person by professors and teaching assistants has been shown to have a negative impact on female students’ grades \cite{cwik2022}. Given that recognition by others has been shown to greatly impact women of color’s science identities \cite{carlone2007}, this lack of recognition by professors and teaching assistants likely impacts their physics identities as well.

A recent research team has begun to investigate women of color and LGBTQ+ (lesbian, gay, bisexual, transgender, and queer) women in physics and how the
intersection of these social identities affects their collegiate experiences \cite{quichocho2020, quichocho2019a, schipull2019}. These works focus primarily on investigating physics identity and have found that some students may fragment or integrate certain parts of their identities into different physics settings, in order to avoid negative outcomes or to be able to interact with certain peers. Furthermore, these findings suggest that being able to reconstruct the stereotype of who does physics to align with their own identities, while also having strong academic and community supports within their physics departments significantly improves the development of a physics identity.

Another recent study conducted an analysis of the experiences of professionals in STEM fields (including physics) utilizing the intersection of gender, race, sexuality, and disability status, making a total of 32 different intersectional groups. This work found that those in the category of white, able-bodied, heterosexual men did experience intersectional privilege providing them with more social inclusion, professional respect, and career opportunities than individuals in the other intersectional categories \cite{cech2022}.

While much work has been conducted over the decades to better understand and improve gender and racial disparities within physics, the majority of the research thus far has focused specifically on women or people of color. Although more recent research has begun investigating the intersection of gender and racial identities and has also begun including sexuality and/or disability as an underrepresented identity, such studies are rare. Furthermore, there have been no research studies that we could find with physics  students  that  also  include traditionally excluded identities (such as international student status, low socioeconomic status, or first-generation college student status) or that explored how the intersection of those identities with other identities affects their experiences. Understanding how having multiple marginalized identities affects students’ experiences in physics is crucial in determining what compounding inequities may arise and in developing a proper course of action to address these issues. Our work detailed in this paper aimed to address this gap in the literature by exploring the challenges of students pursuing an undergraduate physics major, and the impact the intersection of multiple underrepresented identities had on their experiences.  Our research questions for this study were:

\begin{enumerate}
    \item What challenges do undergraduate students face when pursuing a degree in Physics/Astronomy?
    \item Which of these challenges are unique to, or compounded for, individuals with underrepresented identities?
    \item How are these challenges experienced by individuals with specific underrepresented identities or who live at the intersection of multiple underrepresented identities? 
\end{enumerate}

We would like to note that when we utilize the term intersectionality in our work, we are referring to a framework by which one views the human experience through the lens of an interconnected system of social categories to which one belongs \cite{shields2008, rosesalazar}. Although the term was originally coined by Kimberlé Crenshaw \cite{crenshaw1989} to better explain the experiences of Black women in the United States, the concept of intersectionality has since been expanded to explore the intersection of a multitude of social identities on one’s lived experiences.  In our work, we utilize the lens of intersectionality to understand and interpret the experiences of our participants.

Furthermore, when we use the term underrepresented identities in our work, we are acknowledging not only the historical numerical underrepresentation within the field of physics of individuals with such identities but also the power differentials that exist.  These power differentials marginalize the experiences of individuals who claim membership to these underrepresented identities.

\section{Method}
\subsection{Context of the study}
The data for this study were collected at a large public university within the United States. The Physics Department at this institution offers both undergraduate and graduate degrees, including a doctoral degree in physics. Students earning an undergraduate degree from this department are able to choose to either earn a major in physics or a major in physics with a concentration in astronomy.

At the time of the study, the Physics Department at this university had 13 faculty, 12 who present as men and 1 who presents as a woman. Approximately three-quarters of the faculty in the Physics Department are international faculty, all of whom are originally from European, North American, or South American countries.

\subsection{Data collection}
The participants for this study were undergraduate students enrolled as physics majors at this university. To recruit participants for this study, the researchers obtained a list from the university of all 30 undergraduate students who were declared physics majors at the institution at the time of the study. Based on this list, of the 30 undergraduate majors within the department, 9 were labeled as female in the university’s computer system and the remaining 21 were labeled as male\footnote{Note, we recognize that the sex of individuals provided by the computer system is problematic given that it only considers sex in the binary and likely is providing an individual's sex assigned at birth. Therefore, this information may not be accurate to the actual demographics of the undergraduate students within the department.}. Four of the undergraduate majors were labeled as a racial/ethnic minority in the computer system; the remaining 26 were labeled as white. In particular, 18 of the 30 majors were labeled as white males.

In order to ensure a diverse group of participants for this study, an email invitation was sent to all 12 physics majors labeled in the university’s system as either ''female" or a racial minority. The researchers also randomly selected a subset of physics majors who were labeled as white males in the university’s system and sent them an email inviting them to participate in the study as well. In the end, 11 students chose to participate.

For this study, the researchers used a phenomenological case study approach. Through this genre of qualitative research, scholars generally conduct in-depth, iterative interviews using primarily open- ended questions to seek to understand the lived experiences of members of a specific group who have undergone a similar phenomenon \cite{rossman2016, seidman2013}.  The similar phenomenon that the participants in this study were experiencing was that of a physics major at this institution. The data for this study were collected through a series of two in-depth individual interviews with each participant. The data collected during the first interview focused on the students’ pre-college science experiences and how they came to become a physics major. The data collected during the second interview focused on the details of their experience as physics students throughout their college careers. The length of the interviews ranged from 26 minutes to 115 minutes each, with the typical interview lasting approximately 80 minutes. Each interview was audio recorded and transcribed. All participants were given a pseudonym and all identifying information was redacted from the interview transcripts prior to analysis.

In addition to the interviews, participants were asked to complete an online demographic survey \cite{supplement}. Each participant was sent a link to the survey prior to participating in the first interview. This survey was comprehensive and was used to obtain information about the participants’ self-identified underrepresented identities. The underrepresented identity categories that we collected data on were: race/ethnicity, gender, sexuality, international student status, first-generation college student status, and socio-economic status. During the interviews, two of our participants also shared with us that they had a learning and/or physical disability that affected their experiences as well, and therefore we included those as part of their underrepresented identities in our analysis. We acknowledge that there may be more of our participants who have a learning and/or physical disability, however if that is the case, they did not express it as affecting their experiences as a physics major.

Of the 11 participants in our study, four identify as women, five identify as men, and two identify  as  non-binary.  Four  of  the participants identify as a racial/ethnic minority.	Additionally,	other underrepresented identities of our participants (and the frequency of occurrence) include: international student status (1), an underrepresented sexual orientation (4), first-generation college student (5), low socioeconomic status (3), physical disability (1), and learning disability (2). All of the participants were between the ages of 18 and 22. 

Of our 11 total participants, any single participant identifies with between 0 to 5 underrepresented identities (URI’s). In total, we had one participant who identifies with 5 URI’s, two participants who identify with 4 URI’s, two participants who identify with 3 URI’s, three participants who identify with 2 URI’s, one participant who identifies with 1 URI, and two participants who identify with 0 URI’s.  This information is summarized in Table \ref{URI}.

\begin{table}[h]
    \centering
    \begin{tabular}{|c | c|}
    \hline
       Number of URI's   &  Number of Participants \\
       \hline
       5 & 1 \\
       \hline
        4 & 2\\
        \hline
         3 & 2\\
         \hline
         2 & 3\\
         \hline
         1 & 1\\
         \hline
         0 & 2 \\
         \hline
    \end{tabular}
    \caption{Number of URI's with the corresponding number of participants.}
    \label{URI}
\end{table}

\subsection{Data analysis}
The research team, consisting of the three authors of this manuscript, analyzed the anonymized transcripts. Initial codes were developed inductively by the researchers based on themes we identified within the transcripts \cite{corbin2015}. Each interview was then individually coded by at least two of the three researchers on the project. It was agreed that the size of the unit for analysis would be a meaning unit; thus, it could range in length from a few words to a few sentences, as long as the data chunk had a consistent meaning \cite{campbell2013}. Throughout the process, all three researchers would meet regularly to discuss the coding together and reconcile any disagreements within the coding process. During these meetings, we introduced new codes to the codebook as necessary. Once all transcripts were coded, two of the researchers re-coded all of the interview transcripts with the additional codes and reconciled any differences. Finally, once the coding process was completed, all three researchers examined the data instances for each code, identifying themes that address our research questions.

As is the case with all social science research, it is important to acknowledge the bias that may exist within the research process. We acknowledge that we approached this work with a variety of positionalities. One of the members of the research team identifies as a white man and has four underrepresented identities (based on the identity categories that we included in this work). Another member of the research team identifies as a Latina woman and has five underrepresented identities. The final member of the team identifies as a Hispanic woman and has three underrepresented identities. Within the identity categories that we included in this work, at least one of us identifies as a member of an underrepresented group in each category (gender, race/ethnicity, sexuality, international status, first generation college student, low socioeconomic status, and disability).  We believe that our varied identities and lived experiences were a strength of this research team and allowed for a deeper understanding of the phenomena discussed by our participants.

Several measures were taken to ensure trustworthiness of this study. First, the interviews were conducted jointly by two of the authors of this manuscript, both of whom were undergraduate students studying physics at the time of the interviews. Their positions as physics students with varied underrepresented identities awarded them with an insider researcher membership, providing the interviewers with a shared language and experiential base, as well as legitimacy and credibility with the participants \cite{dwyer2009}.

Furthermore, the third researcher of this study is not a member of the physics community. As such, the third researcher is an outsider researcher, which provided the team with an outsider perspective. While we recognize that being an insider researcher does not necessarily affect objectivity, by including an outsider researcher on the team it addresses concerns that some scholars have expressed with the potential of insider researchers having a ``personal position being ‘too close’ for objectivity" \cite{fleming2018}. Furthermore, by conducting two interviews with each participant, this provided extended observations and supplied opportunities for member checking \cite{lincoln1985}.

Additionally, we utilized investigator triangulation, by ensuring each transcript was coded by at least two different members of the research team and having all differing codes reconciled in conjunction with the third member of the team. This intercoder agreement process allowed for multiple checks on coding and interpretation, to ensure stability and accuracy of the coding process \cite{campbell2013}.

\section{Results}
In this section, we begin by discussing the general challenges experienced by participants when pursuing their physics degree, making a point to emphasize when one of their underrepresented identities compounded those experiences. We then discuss the challenges that were specifically unique to certain underrepresented identity groups.

The participant quotes used in this paper are denoted with their pseudonym and the total number of URI’s of the participant who stated the quote (e.g. 3 URI’s indicates the participant had three underrepresented identities).  Due to the small participant pool for this study, we have chosen not to tie the specific URI’s that each participant identifies with to their pseudonym in order to better protect their identities. We further chose to use the pronouns they/them whenever referring to any of our participants, to also help protect the participants’ identities. As a final note, while many of our participants were earning the physics major with a concentration in astronomy offered by the Physics Department, throughout the paper we use only the word physics in place of physics/astronomy for readability purposes.

\subsection{General Challenges}
Although all of the participants experienced some form of support while pursuing the physics major, they all also experienced at least one challenge, regardless of their number of URI’s. While most obstacles were not specific to any single URI, it is important to note that the participants with a higher number of URI’s discussed these challenges more frequently and sometimes within the context of their URI’s. The most prevalent challenges discussed by our participants were: difficult course content, pedagogy, competitive environment, lack of community, and lack of mentorship. In this section we will discuss each of these challenges and will emphasize when certain URI’s compounded the complexity of the experience.

\subsubsection{Difficult Content}
A concern that all of our participants expressed while pursuing their physics degree is the difficult content and excessive mathematics load needed to do physics. This not only involved the depth of the mathematics classes that they had to take, but also how the mathematics was utilized within their physics courses. For example, many of the students talked about having to do a lot of memorization and symbol manipulation for their physics classes. Combining this with the fast-paced environment of the physics classroom and the time-consuming nature of the content, the participants often felt overwhelmed by what they needed to learn to be successful in their physics classes. As Beau  noted:

\emph{Remembering definitions and concepts and formulas, I say sometimes. I mean, also physics is about math. So you need to remember the math… So depending, it kind of gets sometimes confusing because you're learning so many new concepts, you kind of have to develop different, like a way to understand things. It's challenging, it's enjoyable, but it's challenging, and not every time you enjoy it. (Beau, 4 URI’s)}

Beau talked about how one of their physics classes was so much work and so time-consuming, that they actually had to quit their job that semester so they could focus on that one class.

\emph{Yeah, because before I had C-, C+ in the class, and then I was like, you need to focus on this class. I quit my job. And that, I started putting two hours into $\langle$watching$\rangle$ the $\langle$lecture$\rangle$ videos every day… Your grade basically relies on one exam. (Beau, 4 URI’s)}

Notably, Beau identifies with the low socioeconomic URI, and thus had no means of additional financial support when they made this decision to quit their job, putting additional strain on themself in order to be successful in this course.

In addition to the difficult, fast-paced nature of their classes, some participants noted that they also experienced adversities during college due to a lack of knowing how to approach professors and not being familiar with efficient study methods. Brooklyn’s experience, as a first- generation college student, is an example of a compounded experience by those with parents who did not have a post high school education and did not know how to navigate the system.

\emph{I think high school prepared me for more, kind of just get a job, not go to college… I was not prepared to study. I was not prepared to actually read the books given to us. Um, I wasn't really prepared on, like, how to ask professors good questions, and like, how to go to office hours without, like, sending them [the professors] emails. (Brooklyn, 4 URI’s)}
 
Although both first-generation college students and those whose parents have gone to college may experience this difficult transition to college, by not having family members to rely on for guidance and advice, first-generation college students have fewer resources to build on. Furthermore, with the difficult, fast-paced nature of the physics major, our participants talked about how once they got behind in one of their courses it was nearly impossible to catch up.

\subsubsection{Pedagogy and Assessments}
Nine of our 11 participants believed that certain pedagogical choices made by their physics professors ended up hindering their learning within the classroom. For example, some students stated that the amount of content they were expected to learn in their courses often felt unreasonable to learn in one semester. They believed that this created an environment that was too fast-paced, and where it was expected by the faculty that the majority of students would struggle to keep up. Eleanor noted:

\emph{And to throw that much content into one class seems just illegal. Like it's, you're learning everything. You're learning all of pure math in one semester in one class that you have two times a week and then you have to know all of it off the top of your brain for the oral exam. And it just, it feels, it feels wrong. (Eleanor, 1 URI)}

The format of having oral exams in some of their classes was discussed by many of our participants. It is noteworthy that the difficulty of a course with oral exams was discussed by a variety of our participants regardless of their number of URI’s, including Floyd who identifies with 0 URI’s.

\emph{I would say even to this day, $\langle$course$\rangle$ is the hardest physics class and the hardest class I have ever taken in college that demanded a lot; it didn't help that we had oral exams; just like our entire grade dependent on essentially. Uh, yeah, that was, I did learn some cool interesting material from it. Like, I did actually learn the subject material, but [the professor] just went so fast that learning it all was very difficult. And then just the amount that he demanded you to know, like, a moment's notice specifically on the exam was terrifying. (Floyd, 0 URI’s)}

In addition to the fast pace of the courses and oral exams, many of our participants also expressed concerns about not having a variety of assessments in their physics courses. In particular, multiple students stated that they wished they had more homework assigned in their classes. While some participants said they would prefer for this homework to be collected and graded, others thought that it should be assigned but not graded. Either way, many of the participants believed that they did not know how well they were comprehending the material until after taking a test, which was already too late. As Dolly stated:

\emph{I’m also taking $\langle$course$\rangle$. It is a very interesting class. We're not following the textbook too closely. We are kind of jumping around all over the place. There are no homework assignments at all. [The professor] doesn't even assign us problems. He kind of does; he’ll say ``Maybe look at, look at problem 3.5 in the text" or something like that but that's only it in terms of homework. The grades we have in the class are through exams; that's the only feedback we're getting. (Dolly, 2 URI’s)}

The students also often expressed concerns with the culture of exams given within the Physics Department as well. Overall, the students felt like the exams given were not do-able. For the vast majority of their courses, the average on most tests was a failing grade and then the professors would curve the tests accordingly. This set a really negative tone for the students, however, always making them feel discouraged and not smart enough, regardless of what their final grade in the classes ended up being. Eleanor noted:

\emph{It sucks that so many Physics classes, like everyone does bad on exams and everyone said that and I believed them, but then when I started doing bad on exams, it's like, God, this sucks. Like seeing you get a 50$\%$ on an exam. And then knowing that you're at the, like the top of this class, it's like, this is, there's something wrong here. This is not; like something needs to be fixed. And I don't want to go through that again. I don't want to have to see a 50$\%$ on an exam, like I don't, I don't need to see that. (Eleanor, 1 URI)}

As noted by Eleanor, constantly receiving failing grades on exams even when they are ``doing okay" in their classes is really disheartening for the students. These students want more reassurance that their efforts are paying off. As Asher shared:

\emph{The first time I heard ``You are doing okay [in this class]" was from $\langle$Professor D$\rangle$ and that was sophomore year… Please, I want to know I'm doing okay sometimes. I want to not have to go home and throw my things and then fall asleep on the couch because I'm just utterly exhausted. (Asher, 5 URI’s)}

Because of these pedagogical choices and the ``nothing’s ever good enough" culture within the physics classes, it often affected these students in very impactful ways, including their physical health. Asher, who identifies with both a learning and physical disability, discussed experiencing health issues due to one of their courses:

\emph{I really, really struggled with $\langle$Professor E$\rangle$. I was so depressed in that class towards the end. Like that middle portion, not like, really towards the end, but that middle portion. I had a hard time eating, I had a hard time sleeping. Like that, that was when I started having [a specific health problem] actually. And so my doctor is like, ``Hey I don’t know what’s going on, but stop". (Asher, 5 URI’s)}

\subsubsection{Competitive environment}
When discussing the atmosphere of the department, four of the participants stated that they felt like their physics classes were intentionally designed to be competitive and cutthroat, with them having to compete with their peers for grades. Moreover, they also discussed feeling like they needed to compete with their fellow peers in order to get research spots or departmental achievements as well. Asher talked about the pressure they feel to compare themselves to certain students, and how this contributed to competitiveness towards grades and research positions.

\emph{I don’t feel like any of us are doing each other any favors if we’re always comparing ourselves to $\langle$Student A$\rangle$ because that is what I hear from everybody… the way the department has made us feel like, it’s so competitive. (Asher, 5 URI’s)}

This competitive nature made many students feel fixated on their grades. As Eleanor stated:

\emph{I mean, there were some people who were kind of, I don't want to say obsessive over grades, but it was like all they would talk about. And then when you would do worse, it was like you kind of felt bad. (Eleanor, 1 URI)}

Also, Floyd shared that they felt the upper division physics courses were even more competitive than the entry-level ones:

\emph{It did feel more competitive because now you know it wasn't just like an intro level class where you had, like, other majors in there. It was a lot more like this is the physics cohort class so its competitiveness was definitely upped (Floyd, 0 URI’s).}

An additional important detail was that some of the participants talked about a lack of acknowledgement from faculty towards students’ efforts to be successful despite their demanding academic and personal challenges. Asher stated:

\emph{It takes grit to do what we do, it takes a lot of grit. And I know people who have that grit who would never consider doing this. I want us to be recognized for what we’ve been put through and I don’t think the professors do that enough. (Asher, 5 URI’s)}

\subsubsection{Lack of community}
Of our 11 participants, nine of them expressed that throughout their time pursuing the physics major, they felt disconnected from the faculty in the Physics Department. It is important to note that all of these nine participants had at least 1 URI, while the only two participants who had 0 URI’s were the only two who did not express these concerns. Often, the disconnect discussed was expressed as a lack of personal/human interaction with their professors, as well as with the department overall. For instance, Eleanor discussed feeling not cared about by their professors:

\emph{I don't feel like I was encouraged or discouraged. I feel like professors really, I don't want to say they didn't care, but they, it kind of felt like they didn't care, like I don't know. I feel like they always said that they cared. And I think they believe themselves. I don't think they're lying, but I also don't think they actively did things to make me feel like I was cared for at all. (Eleanor, 1 URI)}

For some of the participants with this concern, they believed that this lack of connection actually hindered them academically. As Clio stated:

\emph{I have barely talked to professors outside the class, with the exception of $\langle$course$\rangle$ last semester, I got help with the homework. But other than that, I just don't, because I'm not used to talking to professors or teachers, and I don't know how to do it. So I would rather ask my peers first and then if that doesn't work, just not do it. Which isn’t the way to go about things, but I just don't know how to talk to professors and, like they all kind of scare me just a little bit. (Clio, 3 URI’s)}

Similarly, Dolly shared:

\emph{I don't really communicate too much with [professors]. The biggest time I reached out to them was when I was looking for a research advisor. I reached out to one professor, and I didn't get an answer. I'm assuming it's because he was very busy, and he was very busy, but I didn't really appreciate not getting an answer. (Dolly, 2 URI’s)}

Not receiving a response from this professor made Dolly feel neglected. Although the professor they initially reached out to already had many alternate commitments, they believed that getting any response to their research request, even a no, would have at least not made them feel ignored.

A lack of community amongst the physics students themselves was also a theme that arose within our data. Surprisingly, this was the case for students regardless of whether they were early on in the program or more advanced in their coursework. Some participants credited it partly to a lack of group work in their physics classes, which they believed would allow them to work more closely with their peers to develop those relationships. As Beau noted: 

\emph{I think physics, in the department has been more like a department and not taking the steps to make it like an academic community in a way, like encouragement to make it an academic community. Like the professors would talk about, ``Hey, you have to do a senior research in your senior year"…but they never talk, I mean, they never tell you, ``Hey, how about you group up with your friends?"...they never, like kind of create a community, create a group amongst yourselves, like try to improve that network, the relationship between the students to improve your knowledge. (Beau, 4 URI’s)}

Many of our participants talked about wanting to develop stronger friendships with their physics peers, since they all take so many classes together and have similar professional goals, but they felt like their relationships with those peers generally stayed at surface level, hindering that sense of community. Some of the participants even discussed being unsure if their peers liked them, for example, Asher stated:

\emph{I'm going to be quite honest. I was of the opinion sophomore year that like, all of you hated me. (Asher, 5 URI’s)}

We note that Asher has 5 URI’s and was the only one to express such a strong statement about their peers. In contrast, the two participants with 0 URI’s did not express any concern about fitting in or belonging to their peer group.

\subsubsection{Lack of mentorship}
When asked about their path through the physics major and plans for their future careers, five participants expressed concerns about the lack of mentoring they received concerning these topics, while four discussed not knowing what to do with their degree post-graduation. In particular, we identified a common theme that the students seemed to believe only two paths are available to them as physics majors, as described best by Floyd:

\emph{The implication is that you're either going into grad school to continue that, or you're going to become like an engineer, for instance. (Floyd, 0 URI’s)}

Eleanor shared that they did not know what to do with a physics degree and that they felt like they had to figure it all out on their own:

\emph{I remember sitting at orientation and talking to the advisor and like, realizing that, I had no idea what I was doing. Like I had no idea what careers were out there for physics majors. And I mean, you just kind of went with the flow and hoped that you were doing the right things, and no one ever really went out of their way to tell you like, like, hey, this is what you should be thinking about. (Eleanor, 1 URI)}

Yet another participant, Clio, talked about wanting more guidance on what to do after graduating, in addition to how they should proceed through the physics major itself:

\emph{Mostly [I’d like] just a little bit more guidance with how to work your way through the major and like, specifically toward the end senior project stuff. Like graduation. Guidance toward grad school, if that’s what you want to do. Maybe like better setting of expectations with freshmen of how often should you meet with your department advisor? Just like, give me a handout for me to keep, to hold onto for the four years, so I remember. (Clio, 3 URI’s)}

\subsection{Challenges Unique to Specific Identity Groups}
While all of the participants spoke about challenges that they faced in pursuing a physics major, eight of the nine participants with URI’s mentioned experiencing challenges that were specifically related to one or more of their URI’s. We stress here that throughout the interviews, participants almost always spoke of their URI’s having a negative impact on their experiences in the major. In fact, only two of the participants expressed a benefit from one of their URI’s. To that point, in this section we will focus only on the challenges related to participants’ URI’s, which include being a first-generation college student, coming from a low socio-economic background, having a learning disability, identifying as a woman or non-binary, and being a member of the LGTBQ+ community. Additionally, we place emphasis, when necessary, on those experiences which clearly relate to the intersection of a participant’s URI’s.  We find it worthy to note that while four of our participants identify as a racial minority, none of them discussed challenges that they credited specifically to their race.

Four of our participants discussed the difficulties they experienced by being a first-generation college student. They talked about struggles with navigating college, along with the stress and pressure induced from their family of pursuing a difficult degree. As Dolly stated, their first- generation identity manifested as unawareness in knowing how college works:

\emph{I guess me sort-of being a first-generation college student and not necessarily having the background of how universities work. My mom did a 2-year college degree, and my dad did a trade school or started doing a trade. The only other person in my family so far who has gone to college and completed it was my older cousin and I’m sort of close to her. But she did not do the science realm. So, yeah, I guess for like, not knowing how universities work so much. (Dolly, 2 URI’s)}

In other words, Dolly felt that there were unspoken norms and understandings of how college works that they needed to learn on their own in addition to their academic studies. 

Another participant, Deacon, talked about the stress they felt with being one of the first to pursue college and a science degree. They said:

\emph{So, not only was I the first of the very few in my family to go to the university, I was the only one going into a science field, you know what I mean? So, there was a bit of stress there. Because part of me would just want to make everybody happy, you know, going into the university and I wanted to show people that I could do it. (Deacon, 2 URI’s)}
 
Additionally, Brooklyn discussed the pressure they felt from their family to not just do well at the undergraduate level, but to also pursue a doctoral degree.

\emph{I am a first-generation college student and like I said before, the majority of my family never did anything past high school if they even finished high school. So, I think when people saw that I was the smart one in my family, I was very much pushed to just do more than anyone else has done in my family. Like, even now, my dad has said multiple times like ``A bachelor’s would be really cool, but could you imagine being Dr.$\langle$last name$\rangle$?" (Brooklyn, 4 URI’s)}

As a first-generation college student, having a strong support system is vital for success in college. The lack of information on the inner workings of college and the additional pressure these participants felt from their family, meant that these students had additional challenges to face as compared to their non- first-generation peers.

Some of our participants who were first- generation college students also identified as low socio-economic status, which made their transition to college even more difficult. For example, Carter talked about going to a poorly funded high school that did not have advanced coursework, so they felt they had additional academic challenges when making the transition to college, while simultaneously trying to understand how college works.

\emph{Coming from a school that was, like the highest class we had was like pre-calc. That was a little, a little hard, but then like going into college and realizing, oh, well we still gotta, like, I mean, I guess I could start with calc one, but I definitely should just try and redo a little bit of things. Get used to being in college. So, and I think that's one thing, some people don't like realize making the jump, it was a lot harder. (Carter, 3 URI’s)}

The intersection of this participant’s socioeconomic status and first-generation college student status resulted in them deciding to take additional coursework in college, which was not the case for some of their peers who did not have these intersecting social identities.

Another impactful URI for two of the participants was having a learning disability. The participants with this URI discussed the extra challenges they faced in learning content, as well as physics courses not being a friendly environment for those with a learning disability. Brooklyn talked about how they feel that they are expected to just put up with certain things in physics, despite having a disability:

\emph{In the disability aspect as well, I think it's definitely a conundrum in the science aspect that if you are here, you're expected to be smart, and you're expected to just put up with a lot of things… especially in, like, our Physics Department, there's a lot of things that are kind of, like, oh, can we just like, you know, wave that off? (Brooklyn, 4 URI’s)}

In the context of the interview, when Brooklyn is saying that professors want to ``wave that off", we interpret this as Brooklyn stating that their physics professors did not always want to provide them with all of the accommodations that they are allowed to by law. Additionally, Asher expressed their concerns with the fast-paced nature of physics, and how it is difficult for them to learn in that type of environment.

\emph{Physics and astronomy, the way they're currently taught is they are so fast paced… I learn really slowly, like I have to take my time and. That's because I have… a learning disorder. So it takes me a significantly longer amount of time… just to read a chapter of the book, let alone understand the information and be able to do something with it. So, physics is hard on its own, adding a learning disability on top of it makes it even worse. (Asher, 5 URI’s)}

All of our participants who identified as either a woman or as non-binary expressed that their gender was a challenge for them within the major. These concerns were often related to the lack of non-men in the major and a lack of gender diversity amongst the department faculty. As Eleanor stated:

\emph{I do think that obviously with women there’s less representation. As I said, we don’t have any female professors, no one to look up to as a lady. And I don’t know, like it did come to a point because I mean, obviously a lot of my friends are male just because I was a physics major and most of the friends you can make are men. And there were times where I’d just be like, oh I want some good lady time, like I want some good lady friends. (Eleanor, 1 URI)}

A similar experience was discussed by Clio:

\emph{Like right now in my group of physics majors, it’s me and one other girl so.  And then in the astronomy group there’s a few more.  But it’s just one of those things where it’s like yep, this is still somehow a male dominated field. (Clio, 3 URI’s)}

The obvious lack of female physics instructors  and  peers  at  this  university was a negative experience for many of the women in the program. They believed this lack of representation affected their ability to have mentors and peers who they could relate to. Being a member of the LGBTQ+ community also presented several of our participants with unique challenges. Clio, for example, expressed that they felt like they couldn’t connect with their peers through their LGBTQ+ identity. 

\emph{All of us are ultimately white… so then it’s just like other hidden stuff that’s not race, like queer stuff, but like no one is going to talk about that in class. It was only relatively recently that I had found out that more of the people in the major are queer, and I was like ``What?". And I was like at a party at one of their apartments, and like none of us have mentioned it again. (Clio, 3 URI’s)}

We found it noteworthy that even after these physics majors learned that others in the major were also part of the LGBTQ+ community, they still did not feel comfortable enough to bring it up in later conversations. Based on comments made by multiple participants, it seemed as though such discussions did not feel ``safe" amongst their physics peers. Asher unfortunately felt similarly regarding some of their professors. In particular, they believed one of their professors was intentionally misstating their pronouns.

\emph{Yeah, he has never gotten my pronouns right once. And I’ve told him multiple times at this point. I think it’s on purpose. (Asher; 5 URI’s)}

Based on our data, many of the participants with multiple URI’s felt as though they did not ``fit in" in the field of physics. Asher, in particular, explicitly spoke about the intersection of several of their URI’s and how those make it difficult for them to pursue physics. They also discuss an obvious lack of representation amongst the Physics Department faculty, which was discouraging for them.

\emph{Because [physics is] not a woman-friendly science and back then I was identified as a woman but it’s also not an LGBTQ-friendly science, and it’s really not a disabled friendly science. And so I and I think this is the most important part with this question too is there are not other professors like me in this department. (Asher, 5 URI’s)}

The intersection of Asher’s URI’s left them feeling isolated and uncomfortable, as well as discouraged by the obvious lack of role models amongst the department faculty.

%%%%%%%%%%%%%%%%%%%%%%%%%%%%%%%%%%%%%%%%%%%%%%%%%%%%%%%%%%%%%%%%%%
%%%%%%%%%%%%%%%%%%%%%%%%%%%%%%%%%%%%%%%%%%%%%%%%%%%%%%%%%%%%%%%%%%

\section{Discussion}
This study set out: 1) to explore the challenges faced by college students when earning an undergraduate degree in physics, 2) to learn what challenges were unique to, or compounded for, students with underrepresented identities, and 3) to better understand the challenges experienced by individuals with specific underrepresented identities or who live at the intersection of multiple underrepresented identities.  We will discuss our findings related to these research questions below.

All of our participants faced challenges when pursuing their degree in physics.  We categorized their most common challenges as: difficult content, pedagogy and assessments, competitive environment, lack of community, and lack of mentorship. When considering each of these categories separately, we note that many of these categories involved both academic challenges and environmental challenges within the same category, and that these two types of challenges were not mutually exclusive.  For example, the students who discussed pedagogical and assessment choices made by their professors as a challenge talked about how this challenge affected them academically but also affected the tone of the environment in which they studied and learned.  Therefore, our findings suggest that the general challenges experienced by physics majors are both academic and environmental in nature, and furthermore that these two types of challenges (academic and environmental) are likely to impact each other.

Although we found that all of our participants did face some challenges while pursuing their physics major, it was clear that those with underrepresented identities expressed having more challenges than those without URI’s. In fact, the more URI’s a participant identified with, the more challenges they seemed to encounter. 

Most notable from our results is that only students with underrepresented identities discussed feelings of ``not belonging" amongst their peers and stated there being a lack of community within the department.  Furthermore, the students with underrepresented identities expressed different concerns regarding mentorship.  While participants with underrepresented identities felt that they had inadequate mentoring from their faculty on multiple accounts, those with no underrepresented identities only had concerns with mentoring that surrounded what to do with their degree post-graduation.  Therefore, we conclude that feelings of lack of community and lack of support seem to be more frequently experienced by students with  
underrepresented identities. It is important to note that good mentorship for any student cannot be overstated, especially since previous research suggests that a lack of faculty support can negatively impact the persistence of students in scientific disciplines \cite{ceglie2016}. The role of a mentor is to help cultivate not only the student’s academic abilities, but to also assist them in how to apply for opportunities like fellowships, graduate school programs, jobs, etc. \cite{kram1985, murrell1999}. This assists the students in becoming more confident and well-rounded professionals.  It also helps build community.

It was clear from our data that the participants with one or more URI’s were not receiving the mentorship they desired. While many factors may contribute to this, we wonder if this finding relates in part to the perceptions of faculty in the field regarding ``fixed" vs ``growth" mindsets in physics. It is said that someone has a ``fixed" mindset if they believe that it takes innate talent and aptitude for someone to be successful within a field, while someone is considered to have a ``growth" mindset if they believe that with effort individuals can learn and excel in the field over time \cite{dweck2006}. Previous research has found that faculty, post-doctoral fellows, and graduate students in physics are more likely than scholars in all other STEM fields (other than mathematics), to believe that it takes innate talent to be successful in the field of physics \cite{leslie2015}. These scholars also found a notable correlation that disciplines (both STEM and non-STEM) with strong perspectives of ``fixed" mindsets have lower rates of women and African American scholars in those fields than disciplines with more ``growth" mindsets. In their work, these scholars attribute this finding in part to societal stereotypes of who possesses innate talent in those fields.

Given that our participants with at least one URI do not fit the traditional ``image" of a physicist, it is entirely possible that the faculty at this particular university inadvertently treated them as not having the innate talent that the faculty perceived as needed to be successful in physics, which may have resulted in them receiving fewer mentorship opportunities. In fact, research has shown that underrepresented students who had instructors with ``fixed" mindsets towards their students’ abilities had worse motivational beliefs than students in courses with instructors with ``growth" mindsets \cite{dweck2006}. Furthermore, this ``fixed" mindset may further propagate the harmful stereotype that to be successful in physics you must be brilliant, which can negatively affect those that identify as women or with a disability \cite{meyer2015, traxler2020}.

Additionally, the structure of the content, pedagogical approaches, and competitive environment frequently experienced in the physics courses at this institution caused many of the participants to express feelings of self-doubt and low self-efficacy in their academic abilities, which may also further inhibit one’s connection to physics and the development of a physics identity \cite{hazari2010}. Combining these factors developed an atmosphere that created unnecessary challenges for all of the participants, especially those with URI’s. Indeed, the individualistic nature of the courses do not align with the cooperative learning environment that many of the participants expressed a desire for. This discrepancy may be harmful not only to the students’ abilities to form a collaborative community, but also to their academic development. For many students, particularly those who identify as women, previous research has shown that this lack of community and competitive academic environment can negatively impact their mindset on their ability to do physics and negatively affect their academic performance \cite{kalender2022}.

In addition to participants with underrepresented identities simply having more challenges, we found that there were some challenges that were directly tied to participants with certain underrepresented identities, or that were compounded at the intersection of their URI’s. For example, when our participants with URI’s discussed how their URI’s influenced their challenges, academic challenges were most frequently discussed in the context of having a disability, being a first-generation college student, or being from a low socio-economic status.  In particular, although all of our participants discussed the demanding nature of their physics courses, the participants with a learning and/or physical disability discussed these concerns more frequently.  They found that the pedagogical approaches of many of their physics’ courses presented them with additional academic barriers that they did not necessarily experience in their non-physics courses. Additionally, almost half of our participants were first-generation college students, and each one of them discussed this identity at some point during the interviews. All but one of these participants viewed their first-generation status as providing them with additional academic challenges as a physics major\footnote{The one participant who spoke of this identity as a supprt stated that they used this identity as additional self-motivation to succeed.}.  Furthermore, the participants that had both a first-generation student status and identified as coming from a low socioeconomic background talked about the combination of these identities being a detriment to pursuing their major, an intersectional phenomenon that has been cited elsewhere in the literature \cite{engle2008}. Hence, we believe that the curricular design of the condensed and fast-paced physics coursework is not an accessible environment, especially to those with certain underrepresented identities.

In addition to compounding academic challenges, these specific identities were also found in our study to exacerbate environmental challenges, such as a sense of belonging and access to mentoring from faculty. This phenomenon has been shown by other research as well.  For example, it is known that STEM students with learning disabilities have been underrepresented in the STEM fields due to low learning expectations from others, lack of mentorship and role models, and a lack of individual supports \cite{marino2010, nstc2018}. Research further suggests that first-generation college students, especially those entering STEM fields, are more likely to harbor feelings of self-doubt while simultaneously less likely to feel a sense of belonging. This appears to affect the development of their science identity, which is critical because one’s science identity is a better predictor of academic success for first-generation college students, as compared to their non-first-generation peers \cite{chen2021}. Furthermore, previous research indicates that the intersection of being low-income and first-generation puts such students at a lower rate of attaining their bachelor’s degree, without having an extensive support system \cite{chen2005, choy2000, nunez1998}. Hence, supports within undergraduate physics departments are important to assist students with these identities, particularly because having a strong support system is vital for success in college \cite{gibbons2019}.

A lack of community and support was also noted by our participants in the context of other specific underrepresented identities, such as being a woman or being a member of the LGBTQ+ community. In particular, multiple participants discussed feeling uncomfortable sharing their LGBTQ+ identity with their physics peers and the physics faculty, or discussed feeling as though they were invalidated when they did share.  In this context, these participants found it more difficult to navigate obtaining academic assistance from their physics faculty and felt more isolated from their physics peers.  These experiences and lack of support are consistent with those found through previous research \cite{barthelemy2022}.

When considering the specific underrepresented identities of our participants, one distinction we found important to note was that many of our participants who were members of the LGBTQ+ community or who had disabilities often referred to these as ``hidden identities". In other words, these were identities that they saw as greatly affecting their experiences as individuals and as physics students, yet they were also identities which they could keep hidden from their faculty and peers if they chose to, which they saw as different from their gender and race/ethnicity. In some ways they saw this as giving them more control over their experience but in other ways they saw this as causing them more of a burden, as they did not always feel that they could bring their full selves to the physics classroom.

Finally, we note that our participants who are living at the intersection of multiple underrepresented identities (such as being a member of the LGBTQ+ community and having a disability, or being a member of the LGBTQ+ community and identifying as a woman) discussed it being difficult to navigate multiple underrepresented identities at once and that this compounded their lack of belonging.  Research has shown that these experiences are not unique to our participants and have been reported by other individuals in STEM, where the more underrepresented identities a person has, the more discrimination, isolation, and harassment they have encountered \cite{barthelemy2022, hennessey2019}. Environments like the ones described by members of these groups make it exceedingly difficult for them to feel welcomed and included and make it challenging for them to identify potential allies.

\subsection{Recommendations}
Based on our work and the work of others, it is clear that there is a need to improve the conditions and experiences discussed throughout this paper. We believe there should be a focus on more inclusive pedagogical practices, as well as departmental practices that are inclusive and identify the different needs of students. 

Extant research has shown that utilizing non-lecture teaching techniques within the physics classroom is beneficial for all students, including those with URI’s. Furthermore, research has shown that group-based problem solving can lead to more creativity and inventiveness \cite{bergin2018}. In particular, one approach that has been implemented utilizes problem-based cooperative learning (PBCL) in an undergraduate physics laboratory course, which combines models of problem-based \cite{yoon2012, becerralabra2011} and cooperative \cite{smith2005, heller1992} learning. This method encourages a more learner-centered approach in which the students self-reported a deeper understanding of the scientific process and higher-order problem-solving skills, compared to traditional manual-based laboratory courses. Additionally, the Universal Design for Learning (UDL) \cite{cast_udl}, is a pedagogical approach meant to improve the learning environment for students with special needs by providing multiple means of engagement, representation, and action and expression. Research shows that this approach offers students a non-linear approach to learning \cite{rappolt2012}, assisting those who have a learning disability by focusing on ``fixing" the curriculum rather than repeatedly addressing individual student needs.

Another potential instructional approach is through a method called Modeling Instruction, which includes conceptually based instruction, culturally sensitive instruction, and cooperative learning. This method was studied extensively at a Hispanic-serving institution and reported that the Modeling Instruction method was nearly seven times more successful than standard lecture-based instruction \cite{brewe2010_2}. The results of this study also showed that, by using the Modeling Instruction method, ethnically underrepresented students at this university showed no difference in the odds of success in introductory university physics when compared to the ethnic-majority students. Additionally, female students, when compared to their male counterparts, also showed no difference in the odds of success in introductory university physics.  Therefore, this method was found to provide equitable access to introductory physics content to individuals with a variety of social identities.

We recognize that a dramatic change to physics pedagogy is not a change that can happen without effort. Scholars at the University of Colorado Boulder \cite{bennett2020, hinko2016} have already begun investigating how such changes can be made more easily.  In the context of an informal after-school outreach program, these scholars have created a taxonomy of pedagogical modes displayed by physics educators: instruction mode (lecture style), consultation mode (supporting student engagement without taking control), and participation mode (a student/instructor partner-like approach), as well as have evidence of a mixed-modes method which combines these modes \cite{bennett2020}. This work is aimed at understanding the factors that influence university physics educators’ pedagogy and has implications for constructing university educator training and preparation. In their work, these scholars provide recommendations to improve and broaden formal pedagogical training for university educators in physics.  These recommendations include experiences such as pre- and post-semester training. Although originally intended to train undergraduate physics teaching volunteers to develop strong content knowledge and effective pedagogical skills \cite{chini2016}, we believe similar trainings on inquiry-based teaching methods could be beneficial for physics faculty as well. Since many university-level physics educators have never received formal pedagogical training, aside from potentially minimal training provided to those who had obtained teaching assistant positions during graduate school, we believe pedagogical training opportunities may be interesting and useful to many university physics instructors.

In addition to improving classroom pedagogy, our work suggests that a change in departmental culture may also be beneficial to physics undergraduate students, especially those with underrepresented identities.  A departmental-wide emphasis on creating a welcoming and supportive environment for students entering a physics major would assist in developing a learning environment that helps all students thrive academically and professionally \cite{schipull2019, ceglie2016, atherton2016, thomson2022, brewe2010}.   One way to motivate this change is by providing workshops designed to educate faculty on how curriculum, pedagogy, and departmental practices can be altered to be more inclusive with the intent to ensure a more equitable and welcoming experience for all students. 
 
Additionally, we believe a more inclusive approach to mentoring is also needed. The role of mentorship, especially for those with underrepresented identities, is crucial in helping to develop a student’s physics identity, as well as improving the retention rate and success of minoritized students in physics \cite{kram1985, murrell1999,carterjohnson2016, feder2019}. Having dedicated mentorship programs that focus on delivering an individualized and natural support system to foster a more welcoming and inclusive atmosphere within physics departments is also necessary. 

These types of improvements could make it easier for students to navigate the physics space, break down the long-standing idea of who a physicist is \cite{griffith2010, malcom1976, marchand2013, meador2018, ong2005, wieman2020}, and move towards creating a more inclusive field. For example, recent work on redefining what it means to be a physicist has shown positive impacts on students’ sense of identity and belonging \cite{baylor2022}. Also, there have been recent recommendations from research, which include improving departmental faculty’s understanding of how financial challenges impact learning and engagement, and encouraging more empathy and compassion from faculty for students \cite{pena2022}. Moreover, existing inequities for members of the LGBTQ+ community and others can begin to be addressed by the recent recommendations made by the Ad-Hoc Committee on LGBTQ+ Issues to the American Physical Society \cite{atherton2016}.

\subsection{Limitations and Future Work}
While this work aimed to fill a gap in the literature by including identities not traditionally considered in physics education research, it is not without its limitations. Our sample consisted of only 11 participants at one university. This restricts our ability to draw certain conclusions about physics students’ experience more globally. Furthermore, although we included two participants with 0 URI’s in our work, it is possible that their experiences were not typical of physics students with 0 URI’s. Therefore, our finding that only students with one or more URI’s experience a lack of belonging and lack of community throughout their physics experience may not be generalizable more broadly, depending on the individuals and the context.

Future work should expand on this project by recruiting a larger pool of participants from multiple universities. Furthermore, future research could also be expanded to include students at two-year colleges to address their experiences prior to attending a four-year institution. This approach would allow for a broader and more diverse perspective in order to make more definitive conclusions and suggestions to address the issues presented here.

We also found our participants’ perspectives on ``hidden identities" to be an important one. It would be interesting for future research on intersectional identities to explore, in more depth, the differences in experiences in physics by individuals with such ``hidden" underrepresented identities, individuals with more visible underrepresented identities, and individuals who have a combination of both ``hidden" and more visible underrepresented identities to better understand how having ``hidden" underrepresented identities affect the physics experience differently.

\section*{Acknowledgments}
The authors would like to thank the participants of this study for honestly and openly sharing their experiences with us.  We would also like to thank the anonymous reviewers of this manuscript for their helpful feedback.

%\FloatBarrier

\bibliography{main.bib}

%\printbibliography

\end{document}